\documentstyle[12pt]{article}
\def\beq{\begin{eqnarray}}   \def\eeq{\end{eqnarray}}
\def\t{\theta}

\def\g{gluodynamics}

\begin{document}
\begin{flushright}
NYU-TH/00/05/05 \\
TPI-MINN-16/00\\
UMN-TH-1850/00 \\
hep-th/0007345\\ 
\end{flushright}
\thispagestyle{empty}

\vspace{0.1in}
\begin{center}
\bigskip\bigskip
{\large \bf Vacuum Structure and the Axion Walls in Gluodynamics and
QCD with Light Quarks} 

\vspace{0.5in}      

{Gregory Gabadadze$^{1*}$ and M.~Shifman$^2$}
\vspace{0.1in}

{\baselineskip=14pt \it 
$^1$Department of Physics, New York University,  
New York, NY 10003 } \\
{\baselineskip=14pt \it 
$^2$Theoretical Physics Institute, University of Minnesota, Minneapolis, 
MN 55455} \\
\vspace{0.2in}
\end{center}

\vspace{0.9cm}
\begin{center}
{\bf Abstract}
\end{center} 
\vspace{0.1in}

Large $N_c$  pure gluodynamics was shown to have a set 
of metastable vacua  with the gluonic domain walls interpolating
between them.  The walls may separate the genuine vacuum from an
excited one, or two excited vacua which  are unstable at finite
$N_c$. One may attempt to stabilize them by switching on the axion 
field.
We study how  the  light quarks and  the axion  affect the  structure of
the domain walls. In pure gluodynamics (with the axion field) the axion
walls  acquire a very hard gluonic core.  Thus, we deal with a
wall ``sandwich'' which is stable at finite $N_c$.  In the case of the
minimal axion, the   wall ``sandwich''  is in fact a ``$2\pi$" wall, 
i.e., the
corresponding field configuration interpolates between  identical  
hadronic vacua.  The same properties hold in QCD with three light
quarks and   very large $N_c$.  However,  in the realistic case of 
three-color QCD  the phase corresponding to the axion field profile in
the axion wall is screened by a dynamical phase associated with the
$\eta '$, so that the gluon component of the wall is not excited. We
propose a toy Lagrangian which models these properties and  allows
one to get exact solutions for the domain walls.

\vspace{2cm}
\noindent
$\overline{^*\mbox{\tiny Address after September 1, 2000:
Theoretical Physics Institute, University of Minnesota, Minneapolis, 
MN 55455}}$

\newpage

\section{Introduction}

The early studies \cite{Crewther:1977ce} of the chiral Ward
identities in QCD revealed that the vacuum energy density
depends on the vacuum angle $\theta$ through the ratio 
$\theta /N_f$, where $N_f$ is the number of   quarks with mass 
$m_q\ll\Lambda$. Shortly after, Witten \cite{Witten:1980sp} and 
Di~Vecchia and Veneziano \cite{DiVecchia:1980ve} showed that this
structure occurs  naturally, provided that  there exist  $N_f$ 
states in the theory such that one of them  is the true vacuum,
while others are  local extrema; all are intertwined in the
process of ``the $\theta$ evolution."  Namely, in passage from $\theta
=0$ to  $\theta = 2\pi$, from   $\theta = 2\pi$ to  $\theta = 4\pi$, and
so on, the roles of the above states interchange: one of the local extrema 
becomes the global minimum  and {\em vice  versa}. This would imply,
with necessity, that at $\theta = k\pi$ (where $k$ is an odd integer)
there are two degenerate vacuum states. Such a group of  intertwined
states will be referred to as the ``vacuum family." The crossover at
$\theta = \pi$, $3\pi$, etc. is called the Dashen phenomenon
\cite{DASH}. 

This picture was confirmed by a detailed examination of 
effective  chiral Lagrangians
\cite{Witten:1980sp,DiVecchia:1980ve,Rosenzweig,Arnowitt}  (for a
recent update see \cite{Smilga:1999dh}).  For two and three light 
quarks with equal masses  it was found that the vacuum family 
consists of two or three states respectively; one of them is a global
minimum  of the  potential, while others are local extrema.\footnote{We
stress that the states from the vacuum family need not necessarily lie
at the  minima of the energy functional. As was shown by Smilga
\cite{Smilga:1999dh}, at certain values of $\theta$ some may be
maxima. Those which intersect at $\theta = k\pi$ ($k$ odd) are
certainly the minima at least   in the vicinity of $\theta =k\pi$.}  At
$\theta =\pi$ the levels intersect.  Thus, Crewther's dependence
\cite{Crewther:1977ce}  on  $\theta /N_f$  emerges.

On the other hand, the examination of the effective chiral Lagrangian
with the realistic values of the quark masses,
$m_d/m_u \sim 1.8\,,\,\,\, m_s/ m_d \sim 20$, yields 
\cite{Witten:1980sp,DiVecchia:1980ve,Smilga:1999dh} a drastically
different picture --  the vacuum family disappears (shrinks to one
state); the crossover phenomenon at $\theta=\pi$ is gone as well.

This issue remained in a dormant state for some time. Recently
arguments were given  that the ``quasivacua" (i.e. local minima of the
energy functional), which together with the true vacuum form a
vacuum family,  is an indispensable feature of gluodynamics. The first
argument in favor of this picture  derives \cite{Shifman:1997ua}
from supersymmetric gluodynamics, with supersymmetry softly broken
by a gluino mass term. The same conclusion was reached in Ref.
\cite{Witten:1998uk} based on  a D-brane construction in the limit of
large $N_c$. In fact, one can see that in both approaches 
the number of states  in the vacuum family scales as $N_c$. 
Finally, an additional argument may be found in a cusp
structure which develops once one sums up \cite{Halperin:1998bs}
subleading in $1/N_c$ terms in the effective $\eta '$ Lagrangian. 
At  $N_c =\infty $ the states from the vacuum family are stable,
and so are the domain walls interpolating between them
\cite{Witten:1998uk,ShifmanM}. 

When $N_c <\infty$ the degeneracy and the vacuum  stability
is gone, strictly speaking. It is natural to ask what happens
if one switches on the axion field. 
This generically leads to the formation of the axion domain walls. 
The axion domain wall
\cite{Sikivie:1982qv} presents  an excellent set-up for studying 
the properties of the QCD vacuum under the $\theta$ evolution.
Indeed, inside the axion wall, the axion field (which, in fact, coincides 
with an effective $\theta$) changes  slowly from zero to $2\pi$. 
The characteristic
length scale, determined by the inverse axion mass
$m_a^{-1}$, is huge in the units  of QCD, $\Lambda^{-1}$. 
Therefore, by  visualizing a  set of spatial slices parallel to the axion 
wall,
separated by distances $\gg \Lambda^{-1}$, one obtains
a  chain of  QCD laboratories with  distinct values of $\theta_{\rm
eff}$ slowly varying from one slice to another. In the middle of the wall
$\theta_{\rm eff} =\pi $. 

Intuitively,
it seems clear that in the middle of the axion wall,
the effective value of $\theta_{\rm eff} =\pi$.
Thus, in the central part of the wall
 the hadronic sector is effectively in the regime with
two degenerate vacua, which entails a {\em stable} gluonic wall
as a core of the axion wall. In fact, 
we deal here with an axion wall ``sandwich."
Its core is the so-called D wall, see \cite{GabadShifman}.

Below we will investigate this idea more thoroughly. 
We also address the question  whether this phenomenon
persists  in the theory with  light quarks, i.e., 
in real QCD. Certainly, in the limit $N_c = \infty$
the presence of quarks is unimportant, and the axion wall
will continue to contain the D-wall core. 
As we lower the number of colors, however,  
below  some critical number 
it is inevitable that the regime must change,
the gluonic core must disappear as a result of  the absence of the
crossover.  The parameter governing the change of the regimes
is $\Lambda /N_c$ as compared to the quark mass $m_q$.
At $m_q \ll \Lambda /N_c$, even if one forces the
axion field to form a wall, effectively it is screened by a dynamical
phase whose origin can be traced to the $\eta '$, so that in the 
central part of the axion wall the hadronic sector does not develop
two degenerate vacua. The D walls cannot be accessed in this case
via the axion wall.
 
A part of this paper is of a review character. We collect relevant
assertions scattered in the literature. The main original results --
the occurrence of the D-wall core inside the axion wall in pure
gluodynamics and in QCD with $\Lambda\gg m_q \gg \Lambda /N_c$ --
are presented in Secs. 4 -- 7. 

Recently, the issue of hadronic components of the axion wall in 
the context of a potential with cusps \cite {Halperin:1998bs} 
was addressed in 
\cite{Fugleberg:1999kk,Halperin:1998gx,Forbes:2000gr}.
However, the gluonic component of the axion walls  was  not studied.
The $\eta'$ component in the axion walls was  considered in 
\cite{Evans:1997eq,Fugleberg:1999kk}.  As far as we understand, in 
actuality 
the $\eta'$ component is totally unstable, and cannot be discussed in the
static regime.  

\section{Invisible Axion and Axion Walls}

In this section we briefly  review  the axion set-up, mainly with the
purpose of setting the relevant notation.

The axion was originally  introduced by Weinberg 
\cite{Weinberg:1978ma}
and Wilczek \cite{Wilczek:1978pj} to solve the strong $CP$ problem
which arises  in QCD if physics at very short distances (say,
of order the Planck scale) generates a non-vanishing 
$\t$ term.
In the original version the axion was coupled directly to the light
$u,d$ and $s$ quarks. 

Shortly after, it became clear that
the original construction of Weinberg and Wilczek is not viable from the
phenomenological standpoint  and the axion mechanism was 
further developed:   ``invisible axions'' were introduced.
The concrete version we will keep in mind is 
the KSVZ axion \cite{Kim:1979if,Shifman:1980if} (although other
versions can be considered too \cite{Zhitnitsky:1980tq}). 
One introduces a complex scalar field
$\phi$ coupled to a hypothetical quark field $Q$ in the
fundamental representation of the color SU(3),
with no weak interactions,
\beq
\Delta {\cal L} = \phi \bar Q_R Q_L + \mbox{H.c.}\,.
\eeq
The modulus of $\phi$ is assumed to develop
a large vacuum expectation value $f/\sqrt{2}$, while the argument of 
$\phi$
becomes the axion field $a$, modulo normalization,
\beq
a(x) = f \alpha (x) \,,\quad \alpha (x)\equiv \mbox{Arg} \phi (x)\,,
\qquad f\gg\gg \Lambda\,.
\eeq
Then the low-energy  coupling of the axion to the gluon field is
\beq
\Delta {\cal L} = \frac{1}{f} \, a\, \frac{g^2}{32\pi^2}\,
G_{\mu\nu}^a\tilde G_{\mu\nu}^a\, ,
\label{axglc}
\eeq
so that the QCD Lagrangian depends on the
combination $\theta + \alpha (x)$. 

In general, one could  introduce more than one
fundamental field $Q$, or introduce them  in a higher representation 
of the color group. Then, the axion-gluon coupling (\ref{axglc})
acquires an integer multiplier $N$,
\beq
\Delta {\cal L}' = \frac{1}{f} \, a\, N\, \frac{g^2}{32\pi^2}\,
G_{\mu\nu}^a\tilde G_{\mu\nu}^a\,.
\label{axglcp}
\eeq
This $N$ is sometimes referred to as the axion index,
not to be confused with ${\cal N}$ of extended supersymmetry, nor with 
$N_c$,
the number of colors. The minimal
axion corresponds
to $N=1$. In the general case the QCD Lagrangian  depends on the
combination $\theta + N\alpha (x)$.
The phenomenon of formation of the axion domain walls
is being discussed in the literature for a long time \cite {Sikivie:1982qv}. 
The character of the axion walls depends on $N$. For $N=1$
there is no physical vacuum degeneracy (except at $\theta = \pi$).
Since the
wall  interpolates
between the vacuum and its ``$2\pi$ copy"
it can be bounded by a closed axion string  (see  Ref. \cite{VIL}
for a review).
Thus, such  a wall can
have a finite longitudinal extent. 
This wall is classically unstable as it shrinks its size down
by emitting axions. 
Moreover, the
$N=1$ axion walls are, strictly speaking,
unstable
even if they  have an infinite extent. They
can decay quantum mechanically. The decay process is 
due  to  tunneling between the identical vacua 
separated by a barrier. In fact, a hole can be created in 
the wall -- a domain where the modulus of the field
$\phi \equiv f/\sqrt{2}$ vanishes, and its phase can be ``unwind."
This hole then   expands to infinity removing the wall
completely. Numerically this process is extremely 
suppressed due to the fact that $f$ is very large in the vacuum,
and suppressing $|\phi |$ to make a hole in the wall costs 
a lot of energy.  The suppression factor for tunneling was estimated 
\cite{Sikivie:1982qv} to be 
$\sim {\exp} \left\{ (-{\rm const}\, 
f^2 m_a^{-2})\ln (f^2 m_a^{-2}) \right\}$.
Thus,  the infinite-extent
wall can be considered  stable for all practical purposes.

If $N\geq 2$, there is a residual vacuum degeneracy
of the $Z_N$ type; the walls connecting distinct vacua
{\em must} have infinite area and must be perfectly  topologically stable
(they are cosmologically unacceptable, since  they would over-close 
the Universe \cite {Zeldovich:1974uw}).

Since we have little to add on the process of the wall formation
in the early universe, for our purposes -- consideration of the 
walls in the static environment -- the distinction between
$N=1$ and $N\geq 2$ is unimportant. For simplicity we will deal with 
the $N=1$ axions. All formulae are readily adjustable 
for $N=2$ and higher. 

\section{Two Scenarios (A Signature of the Hadronic Core)}

The invisible axion is very light. Integrating out all other 
degrees of freedom and studying the low-energy 
axion effective Lagrangian must be a good approximation.
The axion effective
potential in QCD can be of two distinct types.

Assuming  that for all values of $\theta$ the QCD vacuum is unique
one  arrives at the axion effective 
Lagrangian of the form 
\beq
{\cal L}_a = f^2
\left[ \frac{1}{2} (\partial_\mu \alpha )^2 +m_a^2
\left(\cos (\alpha  +\theta ) - 1
\right)
\right]\, .
\label{dtwo}
\eeq
The axion potential does  not have to be  
(and generically  is not) a pure cosine; it may have higher
 harmonics. In the general case it is a smooth
periodic  function of 
$\alpha  +\theta $, with the period $2\pi$.
For illustration we presented the potential as a pure cosine.
This does not change the overall picture in the
qualitative aspect. 

As we will see below, a smooth effective potential
of the type (\ref{dtwo}) 
emerges  even if  the (hadronic) vacuum family is non-trivial,
but the transition between the distinct hadronic vacua does not occur
inside the axion wall. This is the case with very light quarks, $m_q\ll
\Lambda /N_c$.  In the opposite limit, one arrives at the
axion potential with cusps, considered below.

In the theory (\ref{dtwo}) one finds 
the axion walls interpolating between the vacuum state
at $\alpha = -\theta$ and the same vacuum state
at $\alpha = -\theta+2\pi$,
\beq
\alpha (z )~+~\theta = 4 \, {\rm arctan}\, ( e^{m_a z})\,,
\eeq
where the wall is assumed to lie in the $xy$ plane, so that
the wall profile depends only on $z$. 
This is the most primitive ``$2\pi $ wall."

The tension of this wall is obviously  of the order of
\beq
T_1 \sim f^2 m_a\, .
\eeq
Taking into account that $ f^2 m_a^2 \sim \chi$ where $\chi$ is
the topological susceptibility of the QCD vacuum, we get
\beq
T_1 \sim \chi /m_a\, .
\eeq
The inverse proportionality to $m_a$ is due to the
fact that the transverse size of the axion wall is very large.

Let us now discuss the axion effective  potential of the second
type.  
In this case the potential has cusps, as is the case in 
pure gluodynamics, where 
the axion effective Lagrangian is of the form
\beq
{\cal L}_a =
 \frac{ f^2}{2} (\partial_\mu \alpha )^2 
+
 \min_\ell \left\{ N_c^2 \Lambda^4
\cos \frac{\alpha +2\pi \ell}{N_c}
\right\}\, ,
\label{axYM}
\eeq
(for a more detailed discussion see below). 
Here the $\theta$ angle was absorbed in the definition of the 
axion field. The axion wall interpolates between
$\alpha = 0$ and $\alpha = 2\pi$. 

What is the origin of this cusp? The cusps reflect a restructuring in the
hadronic sector. 
When one (adiabatically) interpolates in $\alpha$ from $0$ to
$2\pi$ a gluonic order parameter, for instance $\langle G\tilde
G\rangle$,  necessarily experiences  a restructuring 
in the middle of the wall corresponding to the  restructuring 
of heavy gluonic degrees of freedom. In other words,  
one jumps from the hadronic vacuum which initially (at
$\alpha =0$) had $\langle G\tilde G\rangle = 0$ into the vacuum in
which  initially 
$\langle G\tilde G\rangle \neq 0$.
Upon arrival to $\alpha =2\pi$, we find $\langle G\tilde G\rangle =
0$ again. This implies that the central part of such  an axion wall is
dominated  by a 
gluonic  wall. Thus, the cusp at $\alpha = \pi$
generically   indicates the formation of
a hadronic core, the D wall \cite{GabadShifman} in the case at hand.

Returning to the question of the tension
we note that
\beq
\chi &\sim& \Lambda^4 N_c^0\,,\quad  m_a \sim \Lambda^2
N_c^0f^{-1} \,\,\,\mbox{in pure gluodynamics}\,,\nonumber\\[0.2cm]
\chi &\sim& \Lambda^3 N_c m_q\,,\quad m_a \sim
\Lambda^{3/2}m_q^{1/2} N_c^{1/2}f^{-1}
\,\,\,\mbox{in QCD with light quarks}\,,
\eeq
which implies, in turn,
\beq
T_1 \sim \left\{\begin{array}{l}
f\Lambda^2 N_c^0\,\,\,\mbox{in pure gluodynamics}\\[0.3cm]
f\Lambda^{3/2}m_q^{1/2} N_c^{1/2}\,\,\,\mbox{in QCD with light 
quarks}\,.
\end{array}
\right.
\label{dthree}
\eeq
Here $m_q$ is the light quark mass. 

The presence of the large parameter $f$ in $T_1$ makes the axion halo 
the
dominant contributor to the wall tension. The contribution of the
hadronic component contains only hadronic
parameters, although it may have a stronger dependence on $N_c$.
Examining the cusp with 
an appropriately high resolution
 one would observe that it is smoothed on the
hadronic  scale, where the hadronic component of the 
axion wall ``sandwich'' would
become visible. 
The cusp carries a finite contribution to the
wall tension which cannot be calculated in the 
 low-energy approximation \cite{KoganKovnerShifman}.
To this end one needs to consider the hadronic core
explicitly. 
The tension of the core $T_{\rm core} \sim \Lambda^3 N_c$,
while the tension of the axion
halo $T_{\rm halo} \sim f\Lambda^2$ (in pure gluodynamics). 

We pause here to make a comment on the literature.
The consideration of the axion walls in conjunction with 
 hadrons dates back to
the work of Huang and Sikivie, see Ref. \cite{Sikivie:1982qv}.
This work treats the Weinberg-Wilczek $N=2$ axion in 
QCD with two light flavors, which  is replaced by a chiral Lagrangian
for the pions, to the leading order (quadratic in derivatives and
linear in the light quark masses). It is well-known 
\cite{Witten:1980sp,DiVecchia:1980ve,Smilga:1999dh} that
in this theory the crossover phenomenon takes place  at
$m_u=m_d$. In the realistic situation, 
$(m_d-m_u)/(m_d+m_u)\sim 0.3$ considered in Ref. 
\cite{Sikivie:1982qv},
there is no crossover. The pions can be integrated over,
leaving one with an effective Lagrangian for the axion of the type  
(\ref{dtwo}) (with $\alpha \to 2\alpha$). The potential is not pure 
cosine, 
 higher harmonics occur too.
The axion halo exhausts the wall, there is no
hadronic core in this case.

At the same time, Huang and
Sikivie (see Ref. \cite{Sikivie:1982qv}) 
found   an explicit solution for the ``$\pi^0$" component of the 
wall. In fact, this is an illusion. 
The Huang-Sikivie (HS) solution refers to the
{\em  bare}
$\pi^0$ field. To find the physical $\pi^0$ field one must 
diagonalize the mass matrix
at every given value of $\alpha$ (the bare $f\alpha$ 
is the physical  axion field up to
  small corrections $\sim f^2_{\pi}/f^2$
where $f_\pi$ stands for the pion decay constant).  Once this is done, 
one 
observes that the physical pion field,
which is a combination of the bare pion and $f\alpha$, 
is not excited in  the
HS solution. The equation (2.16) in the HS paper  is exactly the condition
of vanishing of the physical pion in the wall profile.
This explains why the wall thickness in the HS work    
is of order $m_a^{-1}$,
with no traces of  the $m_\pi^{-1}$ component.
The crossover of the hadronic vacua at
$\alpha = \pi /2$ (remember, this is $N=2$ model)
could  be recovered in the Huang-Sikivie analysis at
$m_u=m_d$. However, the chiral pion Lagrangian 
predicts  in the two-quark  case the vanishing of the pion mass
in the middle of the wall, for accidental reasons. This is explained in 
detail by A. Smilga, Ref.  \cite{Smilga:1999dh}.

\section{Vacuum Structure in Gluodynamics with  Invisible Axion}
 
First 
 we will summarize arguments in favor of
the existence of a nontrivial vacuum family
in pure gluodynamics.

The first indication 
that the crossover phenomenon
may exist in gluodynamics
comes \cite{Shifman:1997ua}
from supersymmetric Yang-Mills theory,
with supersymmetry being broken by a gluino mass term.
The same conclusion was reached in Ref. \cite{Witten:1998uk}
based on  a D-brane construction in the limit of large $N_c$.
In both approaches 
the number of states  in the vacuum family is $N_c$. 

The Lagrangian of 
softly broken supersymmetric gluodynamics is
\beq
L &=& \frac{1}{g^2}\left\{ -\frac{1}{4} G_{\mu\nu}^a G_{\mu\nu}^a 
+ i\, \bar\lambda^a_{\dot\alpha} 
D^{\dot\alpha\alpha}\lambda^a_{\alpha}
-\left( m \lambda^a_{\alpha}\lambda^{a\alpha}+\mbox{H.c.}\right)
\right\}
\nonumber \\[0.2cm]
& +& 
\theta \, \frac{1}{32\pi^2}\, G_{\mu\nu}^a
\tilde G_{\mu\nu}^a\,,
\label{alsgl}
\eeq
where $m$ is the gluino mass which is assumed to be small, $m\ll 
\Lambda$.

There are $N_c$ distinct chirally asymmetric vacua,
which (in the $m=0$ limit) are labeled by
\beq
\langle \lambda^2\rangle_\ell = N_c\Lambda^3 \exp\left(i \, 
\frac{\theta +2\pi
\ell}{N_c}
\right)\,,\quad \ell = 0,1,..., N_c-1\,.
\eeq
At  $m=0$ there are stable  domain walls interpolating between them
\cite{DvaliShifman}. 
Setting $m\neq 0$  we eliminate the
vacuum degeneracy. 
To  first order in $m$ the vacuum energy density
in this theory is
\beq
{\cal E} = \frac{m}{g^2}\langle \lambda^2\rangle + \mbox{H.c.}
= - mN^2_c\Lambda^3 \, \cos \frac{\theta +2\pi
\ell}{N_c}\,.
\eeq
Degeneracy of the vacua is gone. 
As a result, all the metastable vacua 
will decay very quickly.
Domain walls between them, will be 
moving toward infinity because of the finite 
energy gradient between two adjacent vacua.
Eventually one ends up with a single true vacuum state in the 
whole space.

For each given value of $\theta$ the ground state energy is given by
\beq
{\cal E} (\theta ) = {\rm min}_{\ell}
\left\{ - mN^2_c\Lambda^3 \, \cos \frac{\theta +2\pi
\ell}{N_c}\right\}\,.
\label{dfour}
\eeq
At $\theta = \pi$, $3\pi$, ..., we observe the vacuum degeneracy and
the crossover phenomenon. If there is no phase transition in $m$,
this structure will survive, qualitatively,
even at large $m$ when the gluinos disappear from the spectrum, 
and we recover pure gluodynamics.

Based on a D-brane construction Witten showed \cite{Witten:1998uk}
that in pure SU($N_c$) (non-supersymmetric) gluodynamics 
in the limit $N_c\to\infty$  
a vacuum family does exist:\footnote{This was shown in Ref. 
\cite{Witten:1998uk} assuming that there is no phase transition
in a certain parameter of the corresponding D-brane construction. 
In terms of gauge theory, this assumption amounts of saying that
there is no phase transition as one interpolates 
to the strong coupling constant regime.
Thus, the arguments of \cite{Witten:1998uk}  
have the same disadvantage as those of SUSY gluodynamics
where one had  to assume the absence of the phase transition in
the gluino mass.} the theory has  an infinite group  of states 
(one is the true vacuum,
others are non-degenerate metastable ``vacua'')
which are intertwined  as $\t$ changes by $2\pi\times ({\rm
integer})$, with a crossover at $\t = \pi\times$(odd integer).
The energy density of the $k$-th state from the family is 
\beq
{\cal E}_k(\theta) ~= N_c^2 ~\Lambda^4~
F\left(\frac{\theta + 2\pi k}{N_c}\right)\,,
\label{energ}
\eeq
where $F$ is some $2\pi$-periodic function, and 
the truly stable vacuum for each $\theta$ is obtained  by minimizing 
${\cal E}_k$ with respect to $k$,
\beq
{\cal E}(\theta) ~= N_c^2\,  \Lambda^4\,\mbox{min}_k\,    
F\left(\frac{\theta + 2\pi
k}{N_c}\right)\,,
\label{energg}
\eeq
much in the same way as in Eq. (\ref{dfour}).

At very 
large $N_c$ Eq. (\ref{energg}) takes the form
\begin{eqnarray}
{\cal E}(\theta)~ =~\Lambda^4 ~{\rm min}_k~(\theta + 2\pi k)^2~+ {\cal 
O} \left 
( {1\over N_c}  
\right )~. 
\label{energyN}
\end{eqnarray}
The energy density ${\cal E}(\theta)$ has its
absolute minimum at $\theta =0$. At $N_c =\infty$ the ``vacua"  
belonging to  the vacuum family
are stable but non-degenerate.
To see that the lifetime  of the metastable ``vacuum"
goes to infinity  in
the large $N_c$ limit one can consider
the domain walls which separate these vacua 
 \cite {ShifmanM,GabadC}. These walls are  seen as wrapped 
D branes in the 
construction of \cite {Witten:1998uk}, and they indeed resemble many 
properties of the QCD D branes on which  a QCD string could end. 
We refer to them as D walls because of  their striking similarity to 
D2 branes.  The  consideration of the D walls
has been carried out  \cite {ShifmanM} and leads to the conclusion 
that  the lifetimes of the quasivacua  go to infinity 
as ${\rm exp} (\mbox{const}\, N_c ^4)$.

Moreover, it was argued \cite {DGK,GabadShifman} that the
width of these wall  scales as $1/N_c$ both, in SUSY and pure 
gluodynamics.
To reconcile this observation  with 
the fact that masses of the  glueball mesons scale as $N_c^0$,
we argued  \cite {GabadShifman} that there should exist 
heavy (glue)
states with  masses $\propto N_c $ out of  which the walls are built.
The  D-brane analysis \cite{PStr}, effective Lagrangian arguments
and  analysis of the wall junctions \cite{Gorsky:2000hk},
support this interpretation. These heavy states resemble properties
of the D0 branes. The analogy is striking, as the D0 branes make the
D2 branes from the standpoint of the  M(atrix) theory \cite {BFSS},
so these QCD ``zero-branes'' make the QCD D2 branes (i.e. domain 
walls).\footnote{See also  closely related discussions in Ref.  \cite
{GabadKakush}.} 
The distinct vacua from the vacuum family
differ from each other by a restructuring of these  heavy
degrees of freedom. They
are essentially  decoupled from the glueballs in the large $N_c$ limit. 

Now we switch on the axion
\beq
\Delta{\cal L} = \frac{1}{2}f^2 (\partial_\mu \alpha ) 
(\partial^\mu \alpha ) +\frac{\alpha }{32\pi^2}\, G_{\mu\nu}^a
\tilde G_{\mu\nu}^a\,,
\eeq
with the purpose of studying the axion walls. The potential energy
${\cal E} (\theta)$ in Eq. (\ref{energg}) or (\ref{energyN}) is replaced by
${\cal E} (\theta +\alpha)$. 

Since the hadronic sector exhibits a nontrivial
vacuum family 
and the crossover\footnote{For nonminimal axions,
with $N\geq 2$, the crossover occurs at  $\alpha = k\pi /N$.} 
at $\theta = \pi, 3\pi$,
etc., strictly speaking, it is impossible to integrate out completely the 
hadronic 
degrees of freedom in studying the axion walls.
If we want to resolve the cusp, near the cusp we have to deal with the 
axion field
{\em plus} those hadronic degrees of freedom which restructure. 
In the middle of the wall, at  $\alpha = \pi$,
it is mandatory to jump from one hadronic vacuum to another
-- only then the energy of the overall field configuration
will be minimized and the wall be stable.
Thus, in  gluodynamics the axion wall acquires a D-wall  
core by necessity. 

One can still
integrate out the heavy   degrees of freedom
everywhere except a narrow strip (of
a hadronic size) near the middle of the wall. 
Assume for simplicity that 
there are two states in the hadronic family.  Then the low-energy
effective Lagrangian for the axion field takes the form (\ref {axYM}). 
The domain wall profile will also exhibit
a cusp in the second  derivative. The wall solution takes the form
\beq
\alpha (z ) = 
\left\{\begin{array}{l}
8 \, {\rm arctan}\, \left( e^{m_a z}\,{\rm tan}\frac{\pi}{8}
\right)\,,\quad \mbox{at $z< 0$}\\[0.3cm]
-2\pi +8 \, {\rm arctan}\, \left( e^{m_a z}\,{\rm tan}\frac{3\pi}{8}
\right)\,,\quad \mbox{at $z> 0$}\,,
\end{array}
\right.
\label{dwc}
\eeq
where the wall center is at $z=0$. 

Examining this cusp with 
an appropriately high resolution
 one would observe that it is smoothed on the
hadronic  scale, where the hadronic component of the 
axion wall ``sandwich'' would
become visible. 
The cusp carries a finite contribution to the
wall tension which cannot be calculated in the 
low-energy approximation but can be readily estimated,
$T_{\rm core} \sim \Lambda^3N_c$.

Below we will  examine this core manifestly in a toy solvable model. 
Before doing so, however, we want to elucidate the issue
of the peculiar $N_c$ dependence (or, better to say,
its absence), in Eq. (\ref{energyN}).

\section{Description in Terms of a Three-Index Field}

The  expression for the vacuum energy density (\ref {energyN}) 
seems somewhat puzzling from the point of view of the gluon 
Lagrangian. 
Indeed,   there are $N_c^2-1$ degrees of
freedom in  gluodynamics. Therefore, naively, one expects that the 
vacuum
energy density  in the large $N_c$ limit scales as $\sim N_c^2$.   
However, the leading term in Eq. (\ref {energyN}) scales as $N_c^0$. 
As a possible explanation, one could think of   
a {\it colorless massless} excitation which would give rise to the  
energy density (\ref {energyN}). 
However, there 
are no  {\it physical massless} states  in gluodynamics. 

The explanation to  this apparent puzzle might come
if one introduces
a colorless composite three-index field
which does not propagate any {\it physical} degrees of freedom 
\cite {Aurilia,Lus}.  On the other hand, this field  gives rise \cite
{GabadC} to   precisely the vacuum energy  (\ref {energyN}). 
In a sense, this field is similar to the photon in (1+1)-dimensional QED,
where  a vector particle has no physical  degrees of freedom, but it can
create a constant electric field background which produces  a nonzero
energy.

The  three-index field in gluodynamics  is defined as follows:
\beq
{g^2\over 32\pi^2}~
G_{\mu\nu}^a\tilde G_{\mu\nu}^a ~=~{ 
\varepsilon^{\mu\nu\alpha\beta}~
H_{\mu\nu\alpha\beta} \over 4!}~=~
{ \varepsilon^{\mu\nu\alpha\beta}~
\partial_{[\mu} C_{\nu\alpha\beta]} \over 4!}~, 
\label{HC}
\eeq
where  $H_{\mu\nu\alpha\beta}$ is the field
strength for the  potential $C_{\mu\nu\alpha}$, and the square
brackets denote antisymmetrisation over  all  indices. 
Hence, the $C_{\mu\nu\alpha}$ field can be expressed    
through the gluon fields $A^a_\mu$ as follows:
\beq
C_{\mu\nu\alpha}~=~{1\over 16 \pi^2}(A^a_\mu 
{\overline {\partial}}_\nu
A^a_\alpha-A^a_\nu {\overline {\partial}}_\mu A^a_\alpha-A^a_\alpha 
{\overline {\partial}}_\nu
A^a_\mu+ 2 f_{abc}A^a_\mu A^b_\nu  A^c_\alpha). \label{CA}
\eeq
Here $f_{abc}$ denote the structure constants of the corresponding 
gauge group.  The derivative in this expression is defined as 
$A{\overline {\partial}}B\equiv A (\partial B)-(\partial A) B $.
Note  that the $C_{\nu\alpha\beta}$ field is not a gauge invariant
quantity. If  the  gauge transformation parameter  is denoted as
$\Lambda^a$, the three-index  field transforms as
\beq 
C_{\nu\alpha\beta}\rightarrow
C_{\nu\alpha\beta}+\partial_\nu \Lambda_{\alpha\beta}-
\partial_\alpha \Lambda_{\nu\beta}-\partial_\beta
\Lambda_{\alpha\nu}\,,
\eeq
where $\Lambda_{\alpha\beta}\propto A_\alpha ^a\partial_\beta
\Lambda^a - A_\beta  ^a\partial_\alpha \Lambda^a$.
However, 
the expression for the field strength $H_{\mu\nu\alpha\beta}$ 
is gauge invariant.  

At  energies below $\Lambda$,  all  massive glueballs 
decouple
from the effective Lagrangian of gluodynamics. 
Thus, no physical excitations are left. 
However, there should exist a kinetic term for the  
$C$ field in the low-energy Lagrangian \cite {GabadC}.
This is related to the fact  that the 
correlator of the vacuum topological susceptibility $\chi$ 
at zero momentum
is non-zero in gluodynamics.  
Neglecting all  higher derivative terms and also terms suppressed
in the large $N_c$  limit one arrives at the effective Lagrangian for
the $C$ field of  the form 
\beq
-{1 \over 2\times 4!~\chi}~H_{\mu\nu\alpha\beta}^2~+~\t~
{ \varepsilon^{\mu\nu\alpha\beta}~
H_{\mu\nu\alpha\beta} \over 4!}~+{\cal O}\left ({\partial^2\over 
\Lambda^2},~~
{1\over N_c^2} \right ).
\label{eff}
\eeq  
The first term in this expression reproduces the proper
correlation function 
for the topological susceptibility. 
The second  contribution is just the $\theta$ term.
Once this Lagrangian is set,
it is easy to show that the classical equations of motion 
have a  constant solution
\beq
H_{\mu\nu\alpha\beta}~=~-\chi ~(\t~+~2\pi 
k)~\varepsilon_{\mu\nu\alpha\beta}~,
\label{el}
\eeq 
which reproduces the correct large $N_c$ expression for the energy 
density
(\ref {energyN}). Note that this  
solution persists even if higher derivatives
are included in (\ref {eff}). Moreover, since the $H$ field does not 
propagate the
dynamical degrees of freedom, the large $N_c$ classical solution (\ref 
{el}) 
is also exact quantum-mechanically.  

In this approach, the multiple structure in 
(\ref {energyN}) is related to the quantization of the topological charge
\cite {GabadC}.  
This provides an explanation for the expression  (\ref {energyN}) 
from the point of view of gluodynamics. 

The three-index field $C$ can naturally couple to a D wall.
The corresponding charge of the  D wall is related to the instanton 
number  
in  gluodynamics \cite {GabadC}.
Thus, the D walls are the sources of a constant 
``electric'' field (\ref {el})  
which produces  the vacuum energy density (\ref {energyN}). 

Let us now discuss the mixing of the three-index field
  with the axion, after the latter is switched on. 
At low energies, when all  glueballs are decoupled,
 two new terms emerge in the effective Lagrangian,
\beq
{ 1\over 2}~ (\partial_\mu a )^2 ~+~
{a\over f}~
{ \varepsilon^{\mu\nu\alpha\beta}~
\partial_\mu C_{\nu\alpha\beta} \over 3!}~.
\label{axx}
\eeq
It is known that the pseudoscalar field in four-dimensions 
is dual to 
a two-index antisymmetric gauge field, $B_{\mu\nu}$ \cite 
{Ogievetsky,Kalb}.
That is to say, the axion Lagrangian  (\ref {axx}) can be 
rewritten in terms
of a two-index field. The 
topological charge density, to which the axion is coupled in
(\ref {axx}), is  rewritten in terms of a three-index field 
$C_{\mu\nu\alpha}$
(\ref {HC}). It is intriguing to understand what 
happens with these three- and two-index fields after 
they are coupled  to each other (see also a related  
discussions in \cite {LindeFamily}). 

We can  rewrite (\ref {axx})
in the following equivalent form:
\beq
-{1\over 2} \rho_\mu^2~+~\rho_\mu \partial^\mu a ~+~{a\over f}~
{ \varepsilon^{\mu\nu\alpha\beta}~
\partial_\mu C_{\nu\alpha\beta} \over 3!}~.
\label{rho}
\eeq
Here we have introduced an auxiliary field $\rho_\mu$. 
Equations 
(\ref {axx})  and (\ref {rho}) are equivalent --
to see this one
integrates out $\rho_\mu$ and substitutes 
the result  $\rho_\mu~=~\partial_\mu a$ into 
(\ref {rho}). 

On the other hand, we could first integrate over 
 the axion field in (\ref {rho}). This gives rise to  
the following relation:
\beq
\rho_\mu~=~{1\over f}~\varepsilon_{\mu\nu\alpha\beta}~{
C^{\nu\alpha\beta}~+~\partial^{[\nu} B^{\alpha\beta]} \over 3!}~,
\label{B}
\eeq
where we have introduced an antisymmetric two-index field 
$B_{\alpha\beta}$.
Using the relation (\ref {B}), we find  that 
(\ref {rho}) (or equivalently (\ref {axx})) is proportional to
\beq
{1\over f^2}~\left (  
C_{\nu\alpha\beta}~+~\partial_{[\nu} B_{\alpha\beta]}   \right )^2~.
\label{CB}
\eeq
The sum  in the parenthesis is gauge invariant. In fact, 
it is invariant under both, the Abelian transformations on the $B$ field,
and the non-Abelian transformations of gluons. As we mentioned 
earlier, this latter transformation does not leave 
 $C$  invariant.  The invariance in (\ref {CB}) is restored, however, 
due to compensating transformations of the $B$ field \cite {Nicolai}.

At low energies, when all glueballs are decoupled,
the expression (\ref{CB}) should be combined with the $\theta$ term
and  the    
gauge invariant kinetic term for $C$ given in (\ref {eff}). 
As a result, the expression (\ref{CB}) is nothing but the 
gauge invariant mass term for the three-index field which is a 
superposition of $C$ and $B$ fields. 
In other wards, a mixed state of the $C$ field and the $B$ 
field produces
 a state with the mass  
\beq
m_a^2~=~{\chi\over f^2}~.
\eeq
This is the physical axion 
(similar results were first  obtained in a different context 
in Ref. \cite {LindeFamily} by studying 
correlation functions of the three-index  field.
This is equivalent to
  the effective Lagrangian approach 
adopted here).

Summarizing, we started from gluodynamics
where the $C$ field had no physical components.  The
D walls were the sources of the $C$ field.  
After the axion (represented by $B$)
is switched on, the $C$ field and the
bare axion mix. The mixed three-index field becomes massive and
propagates one massive physical degree of freedom,   the physical
axion. 

The  direct physical consequence 
of this phenomenon is that the  three-form charge of a D wall in a
theory with the axion is {\it screened}. As a result, there will be 
a stationary and stable wall in the theory --  a superposition of 
the axion and
D wall in its core. In the next section we will explicitly find this 
domain wall ``sandwich'' in a toy model. 
  
\section{An Illustrative Model}

To quantitatively describe the axion walls with the D wall core  
one has to  solve QCD, which is way beyond our possibilities.
Our task is more modest. We would like to obtain a qualitative 
description of the axion wall sandwich
which, with luck, can become semi-quantitative. To this end we want to
develop toy models. 
An obvious requirement to any toy model 
is that it must qualitatively reproduce the basic features of
the vacuum structure which we expect in QCD. 
In  SUSY gluodynamics 
it was possible to write down 
a toy model with a ${\bf Z}_{N_c} $ symmetry   
\cite {GabadZ} which ``integrates in"
the heavy degrees of freedom and 
allows one  to 
investigate the BPS domain walls in the 
large $N_c$ limit \cite {DGK} (see also
\cite {DK}).  We will
suggest  a similar model
in (non-supersymmetric) QCD, then switch on axions,
and study the axion domain walls in a semi-realistic setting.
In this model we will  be able to find exact  solutions 
for the D walls and the axion walls.   
	
Here is our a simple toy model which has a proper vacuum 
structure.
If    an appropriate (complex)  glue order parameter is
denoted by $\Phi$, the modulus and phase of this field  
describe  the $0^{++}$ 
and $0^{-+}$ channels of the theory,  respectively.  
The  toy model Lagrangian is
\beq
{\cal L} &=& N_c^2 (\partial_\mu \Phi)^*(\partial_\mu \Phi)
- V(\Phi, \Phi^* )\, , \qquad  V= V_0 + V_1\, ,\nonumber \\[0.2cm]
V_0 &=&  N_c^2 A^2 \left| 1-\Phi^{N_c}e^{-i\theta}\right|^2\,, \nonumber
\\[0.2cm] V_1 &=& \left\{- {\chi N_c ^2\over 2}\,  \Phi \left [ 1+{1\over 
N_c }
(1-
 \Phi^{N_c} e^{-i\theta})  \right] +{\chi N_c ^2\over 2}\right\}
+\mbox{H.c.} \,.
\label{Phi}
\eeq
Here $A$ is a numerical constant of order one,
and $\chi$ is the  vacuum topological susceptibility
in pure gluodynamics (note that $\chi$
is  independent of $N_c $).  The scale parameter $\Lambda$ is
set to unity. 

This model  has the vacuum family composed of
$N_c$ states.  Indeed, the minima of the energy are
determined from the equations
\beq
\left. {\partial V \over \partial \Phi }\right|_{\rm vac}=
\left. {\partial V \over \partial { \Phi^*} }\right |_{\rm vac}=0\,,
\eeq
which have the following solutions:
\beq
\Phi_{\ell \rm vac}={\rm exp} \left ( i\, {\theta+2\pi \ell \over N_c } 
\right )\,,
\quad \ell = 0,1,..., N_c-1\,.
\label{vsol}
\eeq
In the $\ell$-th minimum $V_0$ vanishes, while $V_1$
produces a non-vanishing vacuum energy density,
\beq
{\cal E}_\ell = \chi N_c ^2\left\{
1-\cos\left( {\theta+2\pi \ell \over N_c }\right)\right\} \,.
\eeq
For each given $\theta$ the genuine vacuum is found
by minimization,
\beq
{\cal E} (\theta ) = N_c ^2 \chi  \, \mbox{min}_\ell 
\left\{
1-\cos\left( {\theta+2\pi \ell \over N_c }\right)
\right\}
\,.
\label{cos}
\eeq
The remaining $N_c-1$ minima are quasivacua.
Once the heavy field
$\Phi$ is integrated out, the vacuum energy is 
given by the expression (\ref {cos}); it has cusps at
$\theta =\pi , 3\pi$ and so on.
Needless to say that  the 
potential (\ref {Phi}) has no cusps. 

We will first consider the model (\ref{Phi}) without the axion field,
at $\theta = 0$, in the limit $N_c  = \infty$. In this limit 
the false vacua from the vacuum family  are stable. 

The classical  equation of motion 
defining the wall is 
\beq
N_c ^2\, {\Phi^*}^{\prime\prime}={\partial V\over \partial \Phi}\,,
\label{eq}
\eeq
where primes denote differentiation with respect to $z$
(we  look for a solution which depends
on the $z$ coordinate only).

This is  a differential equation of the  second order.
It is possible, however, to reduce it to a
first order equation. Indeed, Eq. (\ref{eq})
has an obvious ``integral of motion" (``energy"),
\beq
N_c ^2 ~{ \Phi^*}^\prime \Phi^\prime - V = \mbox{Const} = 0\,,
\label{V}
\eeq
where the second equality follows from the boundary conditions.
In the large $N_c $ limit one can  parametrize the field
$\Phi$ as follows ($\rho \sim 1$):
\beq
\Phi\equiv 1+ {\rho\over N_c }\,.
\label{Phirho}
\eeq
Taking the square root of Eq.
(\ref {V}), substituting Eq.
(\ref{Phirho}) and neglecting the terms  of the subleading order in
$1/N_c
$ we arrive at
\beq
{\bar \rho}^\prime =i A N_c \, \left  (1-{\rm exp} \rho\right )\,.
\label{BPS}
\eeq
The phase on the right-hand side can be chosen arbitrarily.
The choice in Eq. (\ref{BPS})
is made in such a way as to make it compatible with the boundary 
conditions
for the wall interpolating between $\Phi_{\rm vac} = 1$ and 
$\Phi_{\rm vac} = \exp (2\pi i / N_c )$.
This is precisely the expression that defines
the domain walls in SUSY \g~ \cite {DGK,DK}. It is not surprising 
that the same equation determines the D walls in non-SUSY \g~--
the fermion-induced  effects are not important for the
D walls in the large $N_c $ limit. 

The solution of this equation was obtained in \cite {DK}.  
In the parametrization $\rho=\sigma+i\tau$
the solution  takes the form:
\beq
&&\cos \tau =(\sigma +1)~{\rm exp}(-\sigma), \nonumber \\[0.3cm]
&&\int_{\sigma(0)}^{\sigma(z)}[{\exp}(2t)-(1+t)^2]^{-1/2}~dt=-
A  N_c |z|\, .
\label{Dsolution}
\eeq
The real part of $\rho$  is   a bell-shaped function with
 an extremum at zero; it vanishes  at $\pm \infty$.
The imaginary part of $\rho$, on the other hand, changes its value from 
$0$ to
$2\pi$.  This determines a D wall  in the large $N_c $ gluodynamics.
The width of the wall scales as $1/N_c $.

The solution presented above is exactly the same as in SUSY 
gluodynamics.
This is not surprising since the
{\em ansatz} (\ref{Phirho}) implies that $V_1$
does not affect the solution --
its impact is subleading in $1/N_c$, while
$V_0$ is exactly the same as in the  SUSY-gluodynamics-inspired model
of Ref. \cite{DGK}. Moreover, for the same reason
the domain wall junctions emerging in this model
will be exactly the same as in  the  SUSY-gluodynamics-inspired model
\cite{GabadShifman}. Inclusion of $V_1$
in the subleading order makes the wall to decay.

Inclusion of the $N=1$  axion field amounts
to the replacement 
$$
\theta \to \theta +\alpha
$$
in Eq. (\ref{Phi}), plus
the axion kinetic term
\beq
{\cal L}_{\rm kin} = \left ({f^2~+~2\Phi^*\Phi \over 2}\right ) 
(\partial_\mu \alpha )^2
+ iN_c (\partial_\mu \alpha ) (\Phi^* \partial_\mu \Phi -
\Phi \partial_\mu \Phi^*)\,.
\eeq 
The occurrence of the mixing between $\alpha$ and the phase of $\Phi$
is necessary, as is readily seen from the softly broken SUSY
 gluodynamics. (To get the potential of the type (\ref{Phi})
in this model, one must eliminate the $G\tilde G$ term by a chiral 
rotation.
Then $m \to m\exp ((\theta +\alpha)/N_c)$ and, additionally
one gets $\partial_\mu\alpha\times$ [the gluino axial current].)
The term $2\Phi^*\Phi $ in the brackets has to be included
to reproduce the correct mass for the axion after the 
physical heavy  state is integrated out. 
The presence of this term signals  that QCD 
dynamics generates not only the potential for the axion but 
also modifies its kinetic term. On the other hand, since
$\Phi^*\Phi\le\Lambda^2$ and, moreover, $\Lambda\ll f$, this 
term can be neglected for all practical purposes.  

We are interested in the configuration with
$\alpha$ interpolating between 0 and $2\pi$.
The phase of $\Phi$ will first adiabatically  follow $\alpha/N_c$,
then at $\alpha \approx\pi$, when the phase of $\Phi$
is close to $\pi /N_c$, it will very quickly 
jump by $-2\pi/N_c$, and then it will continue
to grow  as $\alpha /N_c$, so that when $\alpha$ reaches $2\pi$
the phase of $\Phi$ returns to zero. This jump is continuous, although 
it occurs at
a scale much shorter than $m_a^{-1}$.  This imitates
the D-wall core of the axion wall.
One cannot avoid forming this core, since otherwise
the interpolation would not connect degenerate states --
on one side of the wall we would  have (hadronic) vacuum, on the other 
side an excited state.

In the  large $N_c$ limit one can be somewhat more quantitative.
Indeed, in this approximation the model admits the exact solutions. 
The gluonic core of the wall has the same form as before, Eq.
(\ref {Dsolution}), but the phase $\tau$ is now substituted by
the superposition  $\tau -(\alpha +\theta )$  since  the 
axion field is mixed with the phase of the $\Phi$ field.

This very narrow core is surrounded by a diffused axion halo.
The axion field is described in this halo 
by the solution to the Lagrangian 
(\ref {axYM}). This takes the form:
\beq
\theta~+~\alpha (z)~=~-2\pi~+~4N_c~{\rm arctan}\left 
(e^{m_az}~{\rm tan}{\pi\over 4N_c}
\right )~,~~~~z>0, \nonumber \\
\theta~+~\alpha (z)~=~-
~4N_c~{\rm arctan}\left (e^{-m_az}~{\rm tan}{\pi\over 4N_c}
\right )~,~~~~z<0~.
\label{axionwalls}
\eeq 
Thus, we find explicitly the stable axion wall with a D-wall core. 
Note that this is a usual ``$2\pi$'' wall 
as  it separates two identical hadronic vacua.
As we discussed in the introduction, this wall  is harmless 
cosmologically. 
It  will   be  produced bounded by global axion strings in the early
universe.   Bounded  walls shrink  very quickly 
by decaying  into axions and hadrons.
 
\section{QCD with Three Light Quarks and Axion} 

So far we discussed pure gluodynamics with the axion.
Our final goal is to study QCD  with $N_f= 3$. 
There are two, physically distinct regimes to be considered in this
case. In real QCD 
\beq
m_u,~m_d~\ll~m_s~\sim~{\Lambda\over N_c}~, 
~~~m_u,~m_d,~m_s~\ll~\Lambda~.
\label{I}
\eeq
In this regime the consideration of the chiral Lagrangians \cite
{Witten:1980sp,DiVecchia:1980ve,Smilga:1999dh}, does not exhibit the
vacuum family. We will comment on why
the light quarks screen the vacuum family of the  glue sector,
so that the axion domain wall provides no access to it. 
In the limit (\ref{I}) the effects due to the  
D walls will be marginal. 

On the other hand, in the genuinely  large $N_c$ limit
\beq
{\Lambda\over N_c}~\ll~m_u,~m_d~\ll~m_s~\ll~\Lambda,
\label{II}
\eeq 
physics is rather similar to that of pure gluodynamics. The light quarks
are too heavy to screen the vacuum family of the  glue sector.

In what follows we study the axion walls 
and their hadronic components in the  limits 
(\ref {I}) and (\ref {II}), separately. 

\subsection{One Light Quark}

To warm up, let us start from the theory with one light quark.
In the limit of large $N_c$ this introduces a light meson, ``$\eta '$".
An appropriate effective Lagrangian can be obtained by combining
the vacuum energy density of gluodynamics with what remains from
the Witten-Di Vecchia-Veneziano
Lagrangian at $N_f =1$,
\beq
{\cal L} &=&
 \frac{ F^2}{2} (\partial_\mu \beta)^2 
-V(\beta )\,,\nonumber\\[0.3cm]
V &=& -m_q\Lambda^3N_c \cos\beta +
 \min_\ell \left\{- N_c^2\Lambda^4
\cos \frac{\beta +\theta +2\pi \ell}{N_c} 
\right\}\, .
\label{dsix}
\eeq
Here $\beta$ is the phase of $U \sim \bar q_L q_R$, while  $F^2 \sim
\Lambda^2 N_c$ is the ``$\eta '$'' coupling constant squared. The product
$F\beta$ is the ``$\eta '$'' field. 
The first term in $V$ corresponds to the quark mass term,
${\cal M}U$ + h.c. (see Eq. (7) in Witten's paper \cite{Witten:1980sp}). 
At $N_c =\infty$ the second term in $V$ becomes
$(\beta +\theta )^2$. It corresponds to $(i\mbox{ln det} U +\theta )^2$
in Eq. (11) in  \cite{Witten:1980sp}. The subleading in $1/N_c$ terms
sum up into a $2\pi$ periodic function of the cosine type, with the cusps.
It is unimportant that we used cosine in Eq.  (\ref{dsix}).
Any $2\pi$ periodic function of this type would lead to the same 
conclusions.
The second term in Eq.  (\ref{dsix}) differs from the vacuum energy
density in gluodynamics by the replacement $\theta \to \beta +\theta$.

If $
m_q \ll{\Lambda}/{N_c}$,  the first term in $V$ is
a small perturbation; therefore, 
in the vacuum, $\beta +\theta =2\pi k$, and, hence,
the $\theta$ dependence of the vacuum energy is
\beq
{\cal E}_{\rm vac} (\theta ) = -m_q\Lambda^3N_c \cos\theta\,. 
\eeq
It is smooth,  $2\pi$ periodic and proportional to $m_q$
as it should be on general grounds in the theory with one light quark.

The condition $
m_q \ll{\Lambda}/{N_c}$ precludes us from sending
$N_c\to \infty$.
The would be 
``$2\pi$" wall in the variable $\beta$ is expected to be
unstable. 
 This is due to
the fact that at $N_c \sim 3$ the absolute value of the quark condensate
$\bar \psi \psi$  is not  ``harder" than the phase of the condensate
$\beta$, and the barrier preventing the creation of holes in the
``$2\pi$" wall is practically absent. 

If one closes one's eyes on this instability
one can estimate that the
 tension of the ``$\eta '$" wall is proportional to $\Lambda^3 
N_c^{1/2}$,
with a small correction $m_q\Lambda^2 N_c^{3/2}$ from the quark mass 
term.
The tension of the D-wall core is, as previously, $\Lambda^3N_c$.

In the opposite limit
\beq
m_q \gg\frac{\Lambda}{N_c},\qquad 
\mbox{but $m_q$  still  $ \ll\Lambda$}\, ,
\label{limtwo}
\eeq
the situation is trickier. Now the first term in $V$ is dominant, while the
second is a small perturbation.
 There are $N_c$ distinct vacua in the theory,
\beq
\beta_{\ell}~=~-{2\Lambda \over m_q N_c}~(\t+2\pi\ell)~.
\eeq
Then the $\theta$ dependence of the vacuum energy density
is
\beq
{\cal E}_{\rm vac} (\theta ) = \Lambda^4 \min_\ell ( \theta +2\pi\ell 
)^2\,, 
\label{en2}
\eeq
this is  similar to that in the theory without  light quarks (i.e.,
the same as in 
gluodynamics).
 The ``$\eta '$"   wall is stable at $N_c\to\infty$, with   
a D-wall core in its center. The $\eta'$ wall is a ``$2\pi$''
wall.

From this standpoint, the quark with the mass
(\ref{limtwo}) is already heavy, although
the ``$\eta '$" is still light on the scale of $\Lambda$,
$$
M_{\eta '}\sim m_q^{1/2}\Lambda^{1/2} \ll \Lambda\,.
$$

\vspace{0.2cm}

So far the axion was switched off. What changes if one
includes it in the theory?

\vspace{0.2cm}

The Lagrangian now becomes
\beq
{\cal L} &=&
 \frac{ F^2}{2} (\partial_\mu \beta)^2  +  \frac{ f^2}{2} (\partial_\mu 
\alpha)^2 
-V(\beta ,\alpha )\,,\nonumber\\[0.3cm]
V &=& -m_q\Lambda^3N_c \cos\beta +
 \min_\ell \left\{- N_c^2\Lambda^4
\cos \frac{\beta +\alpha +2\pi \ell}{N_c} 
\right\}\, ,
\label{dseven}
\eeq
where the $\theta$ angle is absorbed in the definition of the
axion field. 

The bare ``$\eta '$" mixes with the bare axion. It is easy to see that
in the limit $
m_q \ll{\Lambda}/{N_c}$
the physical ``$\eta '$" is proportional to $\beta + \alpha$, rather than to
$\beta$. Therefore, even if we force the axion wall to develop,
(i.e. $\alpha$ to evolve from $0$ to $2\pi$)
the ``$\eta '$" wall need not develop. It is energetically expedient to
have $\beta + \alpha=0$.  Thus,  the
effect of  the axion field on the hadronic
sector is totally screened by a dynamical phase $\beta$ 
coming from the quark condensate. In other words,
the axion wall with the lowest tension
corresponds to the frozen physical ``$\eta '$",
 $$\beta + \alpha = 0\,.$$
There is no hadronic core.  The tension of this wall is determined from 
the 
term
$\propto m_q\Lambda^3N_c$. 

[If one wishes, one could add an (unstable)  ``$\eta '$" wall to the axion 
wall.
Then the ``$\eta '$" wall, with the D-wall core will appear
 in the middle of the
axion wall, but they are basically unrelated.
 This will be a secondary 
phenomenon,
and the D wall core will be, in fact, the core of the
``$\eta '$" wall rather than the axion wall.]

If the quark mass is such that (\ref{limtwo})
applies, then
the axion field $\alpha$ cannot be screened,
since we cannot freeze $\beta +\alpha$ 
everywhere in the axion wall profile at zero -- at 
$m_q\gg\Lambda/N_c$,
$\beta$ is proportional to the physical ``$\eta '$" and is much heavier 
than
the axion field. 
Thus,  in this case the axion wall 
will be described by the Lagrangian (\ref {axYM}) and will have a D-wall 
core.
One may also add, on top of it, the
``$\eta '$" wall.
 This will cost 
$m_q^{1/2}\Lambda^{5/2} N_c$ in the wall tension --
still much less than $\Lambda^3 N_c$ of the D-wall
core of the axion wall.

The limit (\ref{limtwo}) is unrealistic. Moreover, in this limit
the D walls taken in isolation, without the
axion walls, are stable by themselves, although they 
interpolate between nondegenerate states \cite{ShifmanM}.

\subsection{Three Light Quarks}

Let us turn to the case of three light flavors. 
The physical picture is quite  similar to that of  
the one-flavor case, see Sec. 7.1.

We assume  the mass matrix ${\cal M}$  in the meson Lagrangian 
to be  diagonal. Therefore, we
will  look  for a diagonal  
$U(3)$ meson matrix which minimizes the potential, 
\beq
U~=~{\rm diag}~\left ( e^{i\phi_1},~~e^{i\phi_2},~~e^{i\phi_3}~\right )~.
\label{U}
\eeq
The potential takes the form
\beq
V ~=~ -\sum_i m_i \Lambda^3 N_c {\rm cos} \phi_i~ +~
 \min_\ell \left \{- N_c^2 \Lambda^4
\cos ~\frac{\sum_i \phi_i +\theta +2\pi \ell}{N_c} 
\right\}~.
\label{V3}
\eeq
As before, we will consider two limiting cases, (\ref {I}) and (\ref {II}).

\vspace{0.2cm}

Let us switch off the axion field first.
In the limit of genuinely light quarks, Eq. (\ref {I}),
when the second term in the potential (\ref{V3})
is dominant, 
 the  solutions
for 
$\phi$'s were found in \cite {Witten:1980sp,DiVecchia:1980ve}. 
They satisfy  the relation  
$\phi_3\simeq 0$ and $\phi_1+\phi_2 =-\t$. 
The corresponding expression for the vacuum energy density is
\beq
{\cal E}_{\rm vac}(\t) ~=~-
N_c~\Lambda^3~\sqrt{~m_u^2+m_d^2+2m_u~m_d\cos \t}~.
\eeq
As in Sec. 7.1, we deal here with a smooth single-valued function of 
$\theta$.
The inclusion of the axion replaces $\theta\to\theta +\alpha\to\alpha$.
The physical $\eta'$ field is 
given by the sum $\sum_i\phi_i+\alpha$. 
It is energetically  favorable to freeze this state. 
Thus, the situation is identical to that in the one-flavor case:
even if the axion wall is forced to develop, the physical $\eta'$
wall (which is now the $\sum_i\phi_i+\alpha$ wall) does not have 
to occur.  The $\eta'$
wall is a gateway to the D wall. In the theory at
hand
 the vacuum angle is screened in the axion wall,
and there is
no D-wall core.

If, nonetheless, the $\eta '$  wall is
formed  due to  some cosmological initial conditions, it will have a D-wall
core (albeit the $\eta '$  wall is unstable in the limit at hand
and cannot be considered in the static approximation).  The 
would-be $\eta '$  wall is independent of the axion wall;
 its effect on the
axion wall  formation is rather irrelevant.  

In addition to this, a ``$2\pi$''
wall could  develop built of nonsinglet mesons, at certain values of the 
quark
masses. There is nothing new we could add to this issue
which is decoupled from the issue of the vacuum family in the glue 
sector
and D walls.

We now pass to the opposite limit  (\ref {II}), when the first term in
 the potential (\ref{V3}) is dominant.
As in Sec. 7.1,
there are $N_c$ distinct vacua with the energy given by (\ref {en2}).
It is straightforward to show that the 
potential for the axion in this case is of the form (\ref {axYM}),
with the cusps which signal the presence of the  D-wall core.  
This is similar to what happens in  gluodynamics. 
One cannot avoid having an  $\eta '$ wall in the middle of the axion wall,
which entails a D wall too.
The D walls  separate the  degenerate vacua. 
Since they ``live" in the middle of the axion wall, they are perfectly 
stable.

(In addition, there can be 
``$2\pi$'' walls in either 
of $\phi$'s or their linear combinations. However, these latter are 
unstable and do not appear in the physical spectrum of 
the theory.)

\section{Conclusions}

Summarizing, we have found that the presence of the axion field
and the axion wall makes the D wall perfectly stable
 in  gluodynamics at finite $N_c$. The D wall develops as a core of the 
axion
wall. It is unavoidable.

In QCD with light quarks the axion wall may or may not generate
the D-wall core. Everything depends on the interplay between the quark
masses,
$\Lambda$ and $N_c$.  
In the realistic case of genuinely light quarks, see Eq.  (\ref {I}),
the phase associated with the axion field in the wall profile
is screened by a dynamical phase (which can be traced back
to the presence of $\eta '$). 
The $\eta '$ is not excited, and neither is the D wall.
There is no D-wall core in the axion wall.

If $N_c$ is increased so that Eq.  (\ref {II}) holds the picture changes 
essentially to that
one deals with in gluodynamics: the $\eta '$ wall is excited,
opening the access
to the D wall.  The  
D-wall core develops in the central part of the axion wall. 
Unfortunately, this limit is unrealistic. 

\vspace{0.2in}

{\bf Acknowledgments}

\vspace{0.2in}

The authors are grateful to B. Bajc, G. Dvali, R. Pisarski, A. Smilga,
and especially A. Zhitnitsky for useful discussions. We thank 
P. Sikivie for useful correspondence.  
The work of G.G. was 
supported by the grant NSF PHY-94-23002. The work of M.S.
was supported by  DOE under the grant number
DE-FG02-94ER408.


\begin{thebibliography}{99}

\bibitem{Crewther:1977ce}
R.~J.~Crewther,
Phys.\ Lett.\  {\bf B70}, 349 (1977);
Acta Phys.\ Austriaca Suppl.\  {\bf 19}, 47 (1978).

\bibitem{Witten:1980sp}
E.~Witten,
Ann.  Phys.\  {\bf 128}, 363 (1980).

\bibitem{DiVecchia:1980ve}
P.~Di Vecchia and G.~Veneziano,
Nucl.\ Phys.\  {\bf B171}, 253 (1980).

\bibitem{DASH}
R. Dashen, 
Phys.\ Rev.\  {\bf D3}, 1879 (1971).

\bibitem{Rosenzweig}
C.~Rosenzweig, J.~Schechter and C.~G.~Trahern,
Phys.\ Rev.\  {\bf D21}, 3388 (1980).

\bibitem{Arnowitt}
R.~Arnowitt and P.~Nath,
Nucl.\ Phys.\  {\bf B209}, 234 (1982).

\bibitem{Smilga:1999dh}
M. Creutz, 
Phys. Rev. {\bf D52}, 2951 (1995) [hep-th/9505112];\\
A.~V.~Smilga,
Phys.\ Rev.\  {\bf D59}, 114021 (1999)
[hep-ph/9805214];\\
M.H.G. Tytgat, 
hep-ph/9909532.

\bibitem{Shifman:1997ua}
M.~Shifman,
Prog.\ Part.\ Nucl.\ Phys.\  {\bf 39}, 1 (1997)
[hep-th/9704114]; \\
N.~Evans, S.~D.~Hsu and M.~Schwetz,
Phys.\ Lett.\  {\bf B404}, 77 (1997)
[hep-th/9703197].

\bibitem{Witten:1998uk}
E.~Witten,
Phys.\ Rev.\ Lett.\  {\bf 81}, 2862 (1998)
[hep-th/9807109].

\bibitem{Halperin:1998bs}
I.~Halperin and A.~Zhitnitsky,
Phys.\ Rev.\  {\bf D58}, 054016 (1998)
[hep-ph/9711398];
Mod.\ Phys.\ Lett.\  {\bf A13}, 1955 (1998)
[hep-ph/9707286];
Nucl.\ Phys.\  {\bf B539}, 166 (1999)
[hep-th/9802095];
Phys.\ Rev.\ Lett.\  {\bf 81}, 4071 (1998)
[hep-ph/9803301];
A.~R.~Zhitnitsky,
Nucl.\ Phys.\ Proc.\ Suppl.\  {\bf 73} (1999) 647.

\bibitem{ShifmanM} 
M.~Shifman,
Phys.\ Rev.\  {\bf D59}, 021501 (1999)
[hep-th/9809184].

\bibitem{Sikivie:1982qv}
P.~Sikivie,
Phys.\ Rev.\ Lett.\  {\bf 48}, 1156 (1982);\\
P.~Sikivie,
Report \#
UF-TP-83-6 (Based on lectures given at 21st Schladming 
Winter School, Schladming, Austria, Feb. 26 -- March 6, 1982); \\
M.~C.~Huang and P.~Sikivie,
Phys.\ Rev.\  {\bf D32}, 1560 (1985);
S.~Chang, C.~Hagmann and P.~Sikivie,
Phys.\ Rev.\  {\bf D59}, 023505 (1999)
[hep-ph/9807374].

\bibitem{GabadShifman} 
G.~Gabadadze and M.~Shifman,
 Phys.\ Rev.\  {\bf D61}, 075014 (2000)
[hep-th/9910050].

\bibitem{Fugleberg:1999kk}
T.~Fugleberg, I.~Halperin and A.~Zhitnitsky,
Phys.\ Rev.\  {\bf D59}, 074023 (1999)
[hep-ph/9808469].

\bibitem{Halperin:1998gx}
I.~Halperin and A.~Zhitnitsky,
Phys.\ Lett.\  {\bf B440}, 77 (1998)
[hep-ph/9807335].
\bibitem{Forbes:2000gr}
M.~M.~Forbes and A.~R.~Zhitnitsky,
hep-ph/0004051.

\bibitem{Evans:1997eq}
N.~Evans, S.~D.~Hsu, A.~Nyffeler and M.~Schwetz,
Nucl.\ Phys.\  {\bf B494}, 200 (1997)
[hep-ph/9608490].
\bibitem{Weinberg:1978ma}
S.~Weinberg,
Phys.\ Rev.\ Lett.\  {\bf 40}, 223 (1978).

\bibitem{Wilczek:1978pj}
F.~Wilczek,
Phys.\ Rev.\ Lett.\  {\bf 40}, 279 (1978).

\bibitem{Kim:1979if}
J.~E.~Kim,
Phys.\ Rev.\ Lett.\  {\bf 43}, 103 (1979).

\bibitem{Shifman:1980if}
M.~A.~Shifman, A.~I.~Vainshtein and V.~I.~Zakharov,
Nucl.\ Phys.\  {\bf B166}, 493 (1980).

\bibitem{Zhitnitsky:1980tq}
A.~R.~Zhitnitsky,
Sov.\ J.\ Nucl.\ Phys.\  {\bf 31}, 260 (1980);\\
M.~Dine, W.~Fischler and M.~Srednicki,
Phys.\ Lett.\  {\bf B104}, 199 (1981).

\bibitem{VIL}
A. Vilenkin and E.P.S. Shellard,
{\em Cosmic Strings and other Topological Defects}
(Cambridge University Press,  1994). 

\bibitem{Zeldovich:1974uw}
Y.~B.~Zeldovich, I.~Y.~Kobzarev and L.~B.~Okun,
Zh.\ Eksp.\ Teor.\ Fiz.\  {\bf 67}, 3 (1974) [Sov. \ Phys. \ JETP {
\bf 40}, 1  (1974)].

\bibitem{KoganKovnerShifman} 
I.~I.~Kogan, A.~Kovner and M.~Shifman,
Phys.\ Rev.\  {\bf D57}, 5195 (1998)
[hep-th/9712046].

\bibitem{DvaliShifman} 
G.~Dvali and M.~Shifman,
Phys.\ Lett.\  {\bf B396}, 64 (1997)
[hep-th/9612128],
(E) {\bf B407}, 4521 (1997). 



\bibitem{GabadC} G. Gabadadze, 
Nucl. Phys. {\bf B552}, 194, (1999) [hep-th/9902191].

\bibitem{DGK} 
G. Dvali, G. Gabadadze, Z. Kakushadze, 
Nucl. Phys. {\bf B562}, 158 (1999) [hep-th/9901032].

\bibitem{PStr}
J.~Polchinski and M.~J.~Strassler,
[hep-th/0003136]; \\
A.~Frey,
hep-th/0007125.

\bibitem{Gorsky:2000hk}
A.~Gorsky and M.~Shifman,
Phys.\ Rev.\  {\bf D61}, 085001 (2000)
[hep-th/9909015].
 
\bibitem{BFSS} T.~Banks, W.~Fischler, S.~H.~Shenker and L.~Susskind,
Phys.\ Rev.\  {\bf D55}, 5112 (1997)
[hep-th/9610043].

\bibitem{GabadKakush} G.~Gabadadze and Z.~Kakushadze,
Mod.\ Phys.\ Lett.\  {\bf A15}, 293 (2000)
[hep-th/9905198]; 
Mod.\ Phys.\ Lett.\  {\bf A14}, 2151 (1999)
[hep-th/9908039].

\bibitem{Aurilia} A.~Aurilia,
Phys.\ Lett.\  {\bf B81}, 203 (1979).

\bibitem{Lus} M.~Luscher,
Phys.\ Lett.\  {\bf B78}, 465 (1978).

\bibitem{Ogievetsky}
V.~I.~Ogievetsky and I.~V.~Polubarinov,
Sov.\ J.\ Nucl.\ Phys.\  {\bf 4} (1967) 156. 

\bibitem{Kalb} M.~Kalb and P.~Ramond,
Phys.\ Rev.\  {\bf D9} (1974) 2273.

\bibitem{LindeFamily} 
R.~Kallosh, A.~Linde, D.~Linde and L.~Susskind,
Phys.\ Rev.\  {\bf D52}, 912 (1995)
[hep-th/9502069].

\bibitem{Nicolai} H.~Nicolai and P.~K.~Townsend,
Phys.\ Lett.\  {\bf B98}, 257 (1981).


\bibitem{GabadZ} G. Gabadadze, 
Nucl.\ Phys.\ {\bf B544}, 650 (1999)
[hep-th/9808005].

\bibitem{DK} G.~Dvali and Z.~Kakushadze,
Nucl.\ Phys.\ {\bf B537}, 297 (1999)
[hep-th/9807140].




\end{thebibliography}
\end{document}